\documentclass[aps]{revtex4}
\usepackage{eurosym}
\usepackage{amsfonts}
\usepackage{amsmath}
\usepackage{amssymb,epsf}
\usepackage{color}
\usepackage{graphicx}
\usepackage{epstopdf}
\usepackage{float}
\usepackage{caption}
\usepackage{subfig}

\begin{document}

\title{Structure of $3D$ gravastars in the context of massive gravity}
\author{H. Barzegar$^{1}$ \footnote{%
email address: h.barzegar@znu.ac.ir}, B. Eslam Panah$^{2,3,4}$ \footnote{%
email address: eslampanah@umz.ac.ir}, G. H. Bordbar$^{5}$ \footnote{%
email address: ghbordbar@shirazu.ac.ir}, and M. Bigdeli$^{1}$ \footnote{%
email address: mbigdeli@znu.ac.ir}}
\affiliation{$^{1}$ Physics Department, College of Sciences, University of Zanjan , Zanjan, Iran\\	
$^{2}$ Sciences Faculty, Department of Physics, University of Mazandaran, P. O. Box 47415-416, Babolsar, Iran\\
$^{3}$ ICRANet-Mazandaran, University of Mazandaran, P. O. Box 47415-416,
Babolsar, Iran\\
$^{4}$ ICRANet, Piazza della Repubblica 10, I-65122 Pescara, Italy\\
$^{5}$ Physics Department, College of Sciences, Shiraz University, Shiraz ,Iran}

\begin{abstract}
In this paper, we investigate a new model of $(2+1)-$dimensional ($3D$)
gravitational vacuum stars (gravastars) with an isotropic matter
distribution anti-de Sitter (AdS) spacetime in the context of massive
gravity. For this purpose, we explore free singularity models with a
specific equation of state. Using Mazur-Mottola's approach, we predict $3D$
gravastars as alternatives to BTZ black holes in massive gravity. We find
analytical solutions to the interior of gravastars free of singularities and
event horizons. For a thin shell containing an ultra-relativistic stiff
fluid, we discuss length, energy, and entropy. In conclusion, the parameter
of massive gravity plays a significant role in predicting the proper length,
energy contents and entropyand parameters of gravastars.
\end{abstract}

\maketitle

\section{Introduction}

Some of the observational pieces of evidence (by the advanced
LIGO/Virgo collaboration) imposed a tight bound on the graviton mass \cite%
{Abbott1,Abbott2}. In addition, there are many theoretical and empirical
limits on the graviton's mass \cite%
{evidmass1,evidmass2,evidmass3,evidmass4,evidmass5}. So one may be motivated
to investigate the effects of massive gravitons on various branches related
to gravitation. On the other hand, the general relativity (GR) is the theory of
a non-trivially interacting massless helicity 2 particles. One of the
interesting modified theories of gravity is related to massive gravity,
which is a modification of GR based on the thought of equipping the graviton
with mass. In theories of massive gravity, the massless helicity 2 particle
of GR becomes massive \cite{Hinterbichler}. In this regard, Fierz and Pauli
first proposed the idea of massive gravitons that are not self-interacting
\cite{1b,2b}. As a result of tests on the Solar system, van Dam, Veltman and
Zakharov (vDVZ) concluded that this original model differed from GR even at
small distance scales \cite{3b,3b1,3b2}. This problem was later solved by
Vainshtein \cite{4b}, who argued that massive gravity could be recovered at
small distances by including nonlinear terms in the field equations. Several
nonlinear completion of massive gravity has shown that this is indeed the
case (see Ref. \cite{5b}). Nonlinear Fierz-Pauli theories, while able to
recover GR through the Vainshtein mechanism, have also revealed another
pathology, the Boulware-Deser ghost \cite{6b}. The ghost problem has only
recently been solved in some papers \cite{7b,8b,10b,11b,12b}. In this
regard, de Rham, Gabadadze and Tolley (dRGT) developed a theory, one of the
interesting ghost-free theories of massive gravity \cite{Hinterbichler,7b}. This
theory uses a reference metric to construct massive terms \cite%
{7b,10b,12b}. These massive terms are inserted in the action to provide
massive gravitons. In 2013, dRGT massive gravity theory was extended by Vegh
\cite{40d}. A ghost-free theory was established by using holographic
principles and a singular reference metric. Many works were done in this
model of massive gravity. Cosmological results, black hole solutions, and
their thermodynamic properties in this massive gravity were investigated by
many authors \cite%
{Mass1,Mass2,Mass3,Mass4,Mass5,Mass6,Mass7,Mass8,Mass9,Mass10,Mass11,Mass12,Mass13,Mass14,Mass15,Mass16,
Mass17,Mass18,Mass19,Mass20}. Also, there are some interesting results of massive
gravity from the astrophysical point of view, for example, the existence of
neutron stars with three times the solar mass \cite{NS1}, and white dwarfs
with masses more than Chandrasekhar's limit \cite{Wht1}. To name a few of
 cosmological points, one can mention: describing the accelerating
expansion of our Universe without requiring any dark energy \cite%
{Cosmass1,Cosmass2}, a suitable description of rotation curves of the Milky
Way, spiral galaxies, and low surface brightness galaxies \cite%
{Panpanich2018}, explaining the current observations related to dark matter
\cite{Babichev1,Babichev2}. From a black hole physics point of view, one can
point out interesting features such as the existence of a remnant for a
black hole which may help to ameliorate the information paradox \cite%
{remnant1,remnant2}, the existence of van der Waals-like behavior in
extended phase space for non-spherical black holes \cite%
{phasemass1,phasemass2}, triple points, and also N-fold reentrant phase
transitions \cite{phasemass3}.

Gravitational vacuum stars (gravastars) are astronomical substances
hypothesized to replace black holes. Gravastars were first proposed by Mazur
and Mottola in Refs. \cite{29f,30f}. This new form of the solution was
introduced as a result of gravitational collapse by expanding the
Bose-Einstein theory. According to this hypothesis, such models contain no
event horizons. By using such structures, we might be able to explain how
dark energy accelerates the expansion of the universe. This could help
explain why some galaxies are more concentrated in dark matter than others
\cite{f}. Visser developed a simple mathematical model for describing the
Mazur-Mottola scenario, and for describing the stability of gravastars by
exploring some realistic values of the equation of state (EoS) parameter
\cite{31f}. Cattoen et al. \cite{32f} extended their results based on the
equations of motion for spherically symmetric spacetime, the anisotropic
factor calculated, and pressure anisotropy analyzed as a factor that can
support relatively high compact gravastars. Carter studied the stability of
gravastar and investigated the existence of thin shells based on the ranges
of parameters involved \cite{33f}. Specifically, he investigated the role of
EoS in the modeling of gravastar structure. Two different theoretical models
for gravastars in an electromagnetic field were presented by Horvat et al.
\cite{34f}. Researchers investigated the effects of electromagnetic fields
on the formulations as well as graphical representations of EoS, the speed
of sound, and the surface redshift. In addition, charged slowly rotating
gravastars were studied by Turimov et al. \cite{38f}.

With a theory of massive gravity, one hopes that the fine-tuning
problem encountered in the cosmological constant problem could get a
technically natural explanation. The argument for technical naturalness is
based on 't Hooft \cite{Hooft}. The general idea is that a small parameter
in a theory is called technically natural if there exists a symmetry that
appears when the value of that parameter is set to zero. In other words,
the principle of naturalness states that is an underlying theory becomes
more symmetric when a parameter involved is set to zero, only then should
this quantity be small in nature. For example, small masses of fermions
(such as electrons) are technically natural because if they were put to
zero, say in the theory of quantum electrodynamics (QED), then chiral
symmetry appears \cite{Dine}. In regard to the extremely small seemingly
fine-tuned value of the bare cosmological constant $\Lambda $, no
such symmetry is known and hence their low values do not conform to 't
Hooft's principle of naturalness. On the other hand, in a theory of massive
gravity with graviton mass $m$ the fine-tuning problem in $\Lambda $ can be redressed into the fine-tuning issue of $\frac{m}{%
M_{pl}}$ ($M_{pl}$ is the Planck mass). When $m$ in a theory is set to zero, it will regain its symmetry under general coordinate invariance, which is the punchline. With a massive graviton, there is hope and sincere motivation that the cosmological constant problem can be solved.

In order to overcome computational and conceptual challenges related to
quantum gravity, one can consider simple models that prevail over these
significant challenges, ideally ones that retain some of the original
conceptual complexity while simplifying the computational process. An
example of such a model is general relativity (GR) in $3D$ spacetime. GR
with dimensions of $(2+1)$ is a good example of such a model. The geometry
of spacetime in $(2+1)-$dimensions has many fundamental similarities with
theories in $(3+1)-$dimensions, which is a great laboratory for many
theoretical ideas. Several fundamental physics issues, including quantum
hall effects, cosmic topologies, parity violations, cosmic strings, and
induced masses have peculiar properties that invite detailed inquiry \cite%
{31s,32s,33s,35s,36s,37s,38s}. Banados, Teitelboim, and Zanelli
at first studied 3D black holes, known as BTZ black holes \cite{1e}. Different
aspects of physics have been impacted by the discovery of BTZ black holes,
such as the thermodynamic properties of these black holes \cite%
{2e,3e,4e,5e,6e,7e} (which contributes to our understanding of gravitational
systems), interactions in lower dimensions \cite{11e}, the existence of
specific relations between BTZ black holes and effective action in string
theory \cite{12e,13e,15e}, and possible existence of gravitational
Aharonov-Bohm effect due to the non-commutative BTZ black holes \cite{16e}.
Additionally, several studies were carried out in the context of AdS/CFT
correspondence \cite{17e,18e,19e}, quantum aspects of 3D gravity,
entanglement, and quantum entropy \cite{23e,24e,25e}. Considering the
importance of 3D spacetime study, in this paper, we will investigate 3D
gravastars as a suitable alternative to BTZ black holes. The existence of
charged gravastars in a $3D$ spacetime is discussed by Rahaman et al. \cite%
{35f}. The researchers examined various physical properties of the charged
gravastars, including length, energy, and entropy. Rahaman et al. \cite{39f}
considered the $3D$ gravastar whose exterior region is elaborated by BTZ
metric. The author discussed various physical features and presented a
non-singular and stable model. Lobo and Garattini \cite{40f} studied
linearized stability analysis with non-commutative geometry of gravastars
and concluded a few exact solutions of gravastars. In Ref. \cite{41f},
Usmani et al. studied a charged gravastar undergoing conformal motion,
examining the dynamics of thin shell formation and the system's entropy.
Barzegar et al. \cite{s} studied AdS $3D$ gravastar in the context of
gravity's rainbow. They extended their results by adding Maxwell's
electromagnetic field and calculated the physical properties of gravastars,
such as proper length, energy, entropy, and binding conditions. The obtained
results show that the physical parameters for the charged and uncharged
states depend significantly on the rainbow functions. Alternatively, it was
shown that classical black holes (such as BTZ black holes) are not possible
in the de Sitter spacetime \cite{39s}. So, in this paper, we will consider
the AdS case for $3D$ gravastars in the context of a modified theory of
gravity, namely dRGT-like massive gravity.

In this paper, we investigate gravastar under spherically symmetric
spacetime with massive gravity. The paper is arranged as follows. Section II
describes the basic framework for massive gravity in $3D$ and their
conservation equation. In section III, we study gravastar structure in the
context of massive gravity in $3D$, and compute the solutions in the three
regions of the gravastar model. In this study, we examine the match between
the interior and exterior regions. Based on the junction conditions, we
compute the gravastar's stability. In the following, we discuss the effects
of massive parameters on the various physical features of gravatar. Finally,
we summarize the results of our investigation.

\section{Field Equations in massive gravity}

The action of massive gravity in $3D$ spacetime with the cosmological
constant $(\Lambda )$ can be written as \cite{Mass15},
\begin{equation}
I=-\frac{1}{16\pi }\int d^{3}x\sqrt{-g}[R-2\Lambda
+m^{2}\sum_{i}^{4}c_{i}U_{i}(g,f)]+I_{matter},  \label{s}
\end{equation}%
where $m$, $R$, $g$ and $f$ are the mass of graviton, the Ricci scalar, the
metric and fixed symmetric tensors, respectively. In Eq. (\ref{s}), $c_{i}$%
's are constants and $U_{i}$'s are the symmetric polynomials of the
eigenvalues of $d\times d$ matrix $\kappa _{\nu }^{\mu }=\sqrt{g^{\mu \alpha
}f_{\alpha \nu }}$ where they can be written in the following form,
\begin{equation}
{\mathit{u}}_{i}=\sum_{y=1}^{i}\left( -1\right) ^{y+1}\frac{\left(
i-1\right) !}{\left( i-y\right) !}{\mathit{u}}_{i-y}\left[ K^{y}\right] ,
\label{U}
\end{equation}%
where ${\mathit{u}}_{i-y}=1$, when $i=y$. It is worthwhile to mention that
in the above relations, the bracket marks indicate the traces in the form; $%
[K]=K_{a}^{a}$ and $[K^{n}]=(K^{n})_{a}^{a}$.

Variation of Eq. (\ref{s}) with respect to the metric tensor, $g_{\mu \nu }$
the equation of motion for massive gravity, leads to (rendering $G=c=1$)
\begin{equation}
G_{\mu \nu }+\Lambda g_{\mu \nu }+m^{2}\chi _{\mu \nu }=8\pi T_{\mu \nu },
\label{G11}
\end{equation}%
where $G_{\mu \nu }$ is the Einstein tensor, $T_{\mu \nu }$ denotes the
energy-momentum tensor, and $\chi _{\mu \nu }$ is the massive term with the
explicit form of the following,
\begin{equation}
\chi _{\mu \nu }=-\sum_{i=1}^{D-2}\frac{c_{i}}{2}\left[ {\mathit{u}}%
_{i}g_{\mu \nu }+\sum_{y=1}^{i}\left( -1\right) ^{y}\frac{i!}{\left(
i-y\right) !}{\mathit{u}}_{i-y}\left[ K_{\mu \nu }^{y}\right] \right] ,
\end{equation}%
where $D$ is related to the dimensions of spacetime. We work on $3D$
spacetime, and so $D=3$. Here, $c_{i}$'s are constants. For $3D$ gravastar,
let us consider a static metric as
\begin{equation}
ds^{2}=f(r)dt^{2}-\frac{dr^{2}}{g(r)}-r^{2}d\theta ^{2}  \label{metric}
\end{equation}%
where $f(r)$ and $g(r)$ are unknown metric functions of the radial
coordinate. An exact solution of the metric (\ref{metric}) can be obtained
by choosing a reference metric as given by
\begin{equation}
f_{\mu \nu }=diag(0,0,-C^{2}),  \label{f}
\end{equation}%
in which $C$ is a positive constant. Considering the metric ansatz (\ref{f}%
), $U_{i}$'s can easily be computed as $U_{1}=C/r$, and $U_{2}=U_{3}=U_{4}=0$%
, which indicates that the contribution of massive gravity in $3D$ spacetime
is arising only from the $U_{1}$.

We assume that the matter distribution in the interior of the gravastar is a
perfect fluid type, given by
\begin{equation}
T_{\mu \nu }=(\rho +p)u_{\mu }u_{\nu }-pg_{\mu \nu },  \label{T1}
\end{equation}%
where $\rho $ represents the energy density, $p$ is the isotropic pressure,
and $u^{i}$ are the components of velocity of the fluid. Using the spacetime
described by the metric (\ref{metric}) together with the energy-momentum
tensor given in Eq. (\ref{T1}), we can obtain the nonzero components of
field equation (\ref{G11}) as
\begin{eqnarray}
\frac{g^{\prime }}{2r}+\frac{m^{2}c_{1}C}{2r} &=&\Lambda -8\pi \rho ,
\label{G1} \\
&&  \notag \\
\frac{gf^{\prime }}{2fr}+\frac{m^{2}c_{1}C}{2r} &=&\Lambda +8\pi p,
\label{G2} \\
&&  \notag \\
\frac{g^{\prime }f^{\prime }}{4f}+\frac{gf^{\prime \prime }}{2f}-\frac{g{%
f^{\prime }}^{2}}{4f^{2}} &=&\Lambda +8\pi p,  \label{G3}
\end{eqnarray}%
where the prime and double prime are representing the first and second
derivatives with respect to $r$, respectively. Combining Eqs. (\ref{G1})-(%
\ref{G3}), we get
\begin{equation}
p^{\prime }+(\rho +p)\frac{f^{\prime }}{2f}=0,  \label{TOV}
\end{equation}%
which is the conservation equation in $3D$ spacetime.

\section{$3D$ Gravastars Structure}

The gravastars can be described with the help of three different zones, in
which zone I is the interior region $(0<r<r_{1})$, zone II is the
intermediate thin shell, with $r_{1}<r<r_{2}$, while zone III is an exterior
region $(r_{2}<r)$. In zone I, the isotropic pressure produces a force of
repulsion over the intermediate thin shell, which is equal to $-\rho $
(where $\rho $ is the energy density). This intermediate thin shell is
supposed to be supported by fluid pressure and ultra-relativistic plasma ($%
p=\rho $). However, zone III can be represented by the vacuum solution of
the field equations. The pressure has zero value in this zone. It contains a
thermodynamically stable solution and maximum entropy under small
fluctuations \cite{29f, 30f}. In this section, we derive the equations of $%
3D $ gravastar field of massive gravity for different regions and analyze
them.

\subsection{Interior Spacetime}

The interior region $(0<r<r_{1}=R)$ of the gravastar follows the EoS $%
p=-\rho $. Hence by using the result given in Eq. (\ref{TOV}), we obtain the
following interior,
\begin{equation}
\rho =constant=\rho _{v},
\end{equation}%
and
\begin{equation}
p=-\rho _{v}.
\end{equation}
By using the Eq. (\ref{G1}), one can get the solutions for $g(r)$ and $f(r)$
from the field equations as
\begin{equation}
g(r)=f(r)=A+\Lambda r^{2}-8M(r)-m^{2}c_{1}Cr,  \label{g(r)}
\end{equation}%
where $A$ is an integration constant and $M(r)=\int {2\pi \rho rdr}$. From
Eq. (\ref{g(r)}), we arrived at the important conclusion that the spacetime
metric thus obtained is a singularity free solution of the gravastars at the
centre. Hence, the active gravitational mass $M(r)$ can be expressed at once
in the following form,
\begin{equation}
M(R)=\int_{0}^{R}2\pi r\rho dr=\pi R^{2}\rho _{v}.  \label{M}
\end{equation}
Here, we note that for the interior region, the physical parameters, viz.
density, pressure and gravitational mass in no way are dependent on the
massive parameter $(m^{2}c_{1})$. We also observe that the quantities $g(r)$
and $f(r)$ depend on the massive parameter $(m^{2}c_{1})$.

\subsection{Intermediate Thin Shell}

It is very difficult to solve the field equations within the non-vacuum
region, i.e., within the shell. However, one can obtain an analytic solution
within the framework of thin shell limit, $0<g(r)\equiv h<<1$. The advantage
of using this thin shell limit is that in this limit we can set $h$ to be
zero to the leading order. Then the field equations (\ref{G1})-(\ref{G3}),
with $p=\rho $, may be recast in the forms
\begin{eqnarray}
h^{\prime } &=&4\Lambda r-2m^{2}c_{1}C,  \label{h} \\
&&  \notag \\
\frac{f^{\prime }}{f} &=&\frac{2m^{2}c_{1}C}{rh^{\prime }}.
\end{eqnarray}

Integrating Eq. (\ref{h}) immediately yields
\begin{equation}
h=g(r)=B+2\Lambda r^{2}-2m^{2}c_{1}Cr,
\end{equation}%
where $B$ is an integration constant. So the other function is
\begin{equation}
f(r)=F_{0}\left( 2\Lambda -\frac{m^{2}c_{1}C}{r}\right) ,
\end{equation}%
where $F_{0}$ is an integration constant. Also, from the conservation Eq. (%
\ref{TOV}) and using the EOS $p=\rho $, one may get
\begin{equation}
p=P_{0}\left( {2\Lambda -\frac{m^{2}c_{1}C}{r}}\right) ^{-1},
\end{equation}%
where $p_{0}$ being an integration constant.

\subsection{Exterior Region}

The vacuum exterior region EoS is given by $p=\rho =0$. Solution corresponds
to a static BTZ black hole in massive gravity is written in the following
form as \cite{Mass15}
\begin{equation}
f(r)=g(r)=\Lambda r^{2}-m_{0}-m^{2}c_{1}Cr,  \label{E2}
\end{equation}%
the parameter $m_{0}$ is an integration constant related to the total mass
of black holes.

\subsection{Junction Condition}

The conditions for matching interior and exterior geometry were introduced
by Darmois \cite{56f} and Israel \cite{57f}. The metric coefficients are
continuous at the junction surface, i.e., their derivatives might not be
continuous at interior surfaces. The second fundamental forms associated
with the two sides of the shell are given in the literature \cite%
{24T,25T,26T,27T,28T,29T}. The surface tension and surface stress energy of
the joining surface $S$ may be resolved from the discontinuity of the
extrinsic curvature of $S$ at $r=R$. The field equation of intrinsic surface
is defined by Lanczos equation \cite{118s} as
\begin{equation}
S_{\alpha \beta }=-\frac{1}{8\pi }\left( k_{\alpha \beta }-\delta _{\alpha
\beta }k_{\gamma \gamma }\right) .  \label{S1}
\end{equation}%
Here $S_{\alpha \beta }$ is the stress-energy tensor for surface, $%
k_{ij}=K_{ij}^{+}-K_{ij}^{-}$ tells the extrinsic curvatures or second
fundamental form, and ($+$) sign indicates the interior surface while ($-$)
sign indicates the exterior surface. The second fundamental form connects
the interior and exterior surfaces of the thin shell are defined as follows,
\begin{equation}
K_{ij}^{\pm }=\left[ -n_{\nu }^{\pm }\left( \frac{\partial ^{2}x_{\nu }}{%
\partial \xi ^{i}\partial \xi ^{j}}+\Gamma _{\alpha \beta }^{\nu }\frac{%
\partial x^{\alpha }}{\partial \xi ^{i}}\frac{\partial x^{\beta }}{\partial
\xi ^{j}}\right) \right] ,  \label{S2}
\end{equation}%
where $\xi ^{i}$s represent the intrinsic coordinates on the shell, and $%
-n_{\nu }^{\pm }$s are the unit normal vectors on the surface of gravastar
in the following form,
\begin{equation}
n_{\nu }^{\pm }=\pm \left( {g^{\alpha \beta }\frac{\partial f}{\partial
x^{\alpha }}\frac{\partial f}{\partial x^{\beta }}}\right) ^{\frac{-1}{2}}%
\frac{\partial f}{\partial x^{\nu }}.  \label{S3}
\end{equation}%
In above equation, $n^{\nu }n_{\nu }=1$, and $f(r)$ illustrates the
coordinate of exterior metric. Surface tension and surface stress of the
junction surface are determined by the discontinuity in the extrinsic
curvature. Now, from Eq. (\ref{S2}) and Lanczos equation in $3D$ spacetime,
we can get surface energy density $(\varphi )$ and surface pressure $(\psi )$
as
\begin{eqnarray}
\varphi &=&\frac{-k_{\phi }^{\phi }}{8\pi },  \label{eqrho} \\
&&  \notag \\
\psi &=&\frac{-k_{\tau }^{\tau }}{8\pi },  \label{eqP}
\end{eqnarray}%
where $\varphi $ and $\psi $ are line energy density and line pressure of $%
3D $ gravastar in massive gravity, respectively. So, according to the
general formalism for $3D$ spacetime \cite{118s} and employing relevant
information into equations (\ref{eqrho})-(\ref{eqP}), and also by setting $%
r=R$, we obtain
\begin{eqnarray}
\varphi (R) &=&\frac{\sqrt{\frac{A}{R^{2}}+\Lambda -8\pi \rho _{v}-\frac{%
m^{2}c_{1}C}{R}}}{8\pi }  \notag \\
&&-\frac{\sqrt{\Lambda -\frac{m_{0}}{R^{2}}-\frac{m^{2}c_{1}C}{R}}}{8\pi },
\label{eqrho2} \\
&&  \notag \\
\psi (R) &=&\frac{\Lambda R-\frac{m^{2}c_{1}C}{2}}{8\pi \sqrt{\Lambda
R^{2}-m_{0}-m^{2}c_{1}CR}}  \notag \\
&&-\frac{(\Lambda R-8\pi \rho _{v}R-\frac{m^{2}c_{1}C}{2}}{8\pi \sqrt{%
A+\Lambda R^{2}-8\pi \rho _{v}R^{2}-m^{2}c_{1}CR}}.  \label{eqP2}
\end{eqnarray}

In the following, we can study the equation of state parameter and stability
of gravastars by using line energy density and line pressure of $3D$
gravastar in massive gravity.

\subsubsection{Equation of State}

At a particular radius $r=R$, the equation of state parameter can be
expressed as follows,
\begin{equation}
\omega =\frac{\psi (R)}{\varphi (R)}.  \label{omega}
\end{equation}%
By using Eqs. (\ref{eqrho2}) and (\ref{eqP2}) in the expression (\ref{omega}%
), the EoS parameter at $r=R$ can be obtained in the following form,
\begin{equation}
\omega =\frac{\frac{\Lambda -8\pi \rho _{c}-\frac{m^{2}c_{1}C}{2R}}{\sqrt{%
\frac{A}{R^{2}}+\Lambda -8\pi \rho _{v}-\frac{m^{2}c_{1}C}{R}}}-\frac{%
\Lambda -\frac{m^{2}c_{1}C}{2R}}{\sqrt{\Lambda -\frac{m_{0}}{R^{2}}-\frac{%
m^{2}c_{1}C}{R}}}}{\sqrt{\Lambda -\frac{m_{0}}{R^{2}}+\frac{m^{2}c_{1}C}{R}}-%
\sqrt{\frac{A}{R^{2}}+\Lambda -8\pi \rho _{v}-\frac{m^{2}c_{1}C}{R}}}.
\end{equation}

\subsubsection{Stability}

\begin{figure}[tbp]
\centering
\includegraphics[scale=0.3]{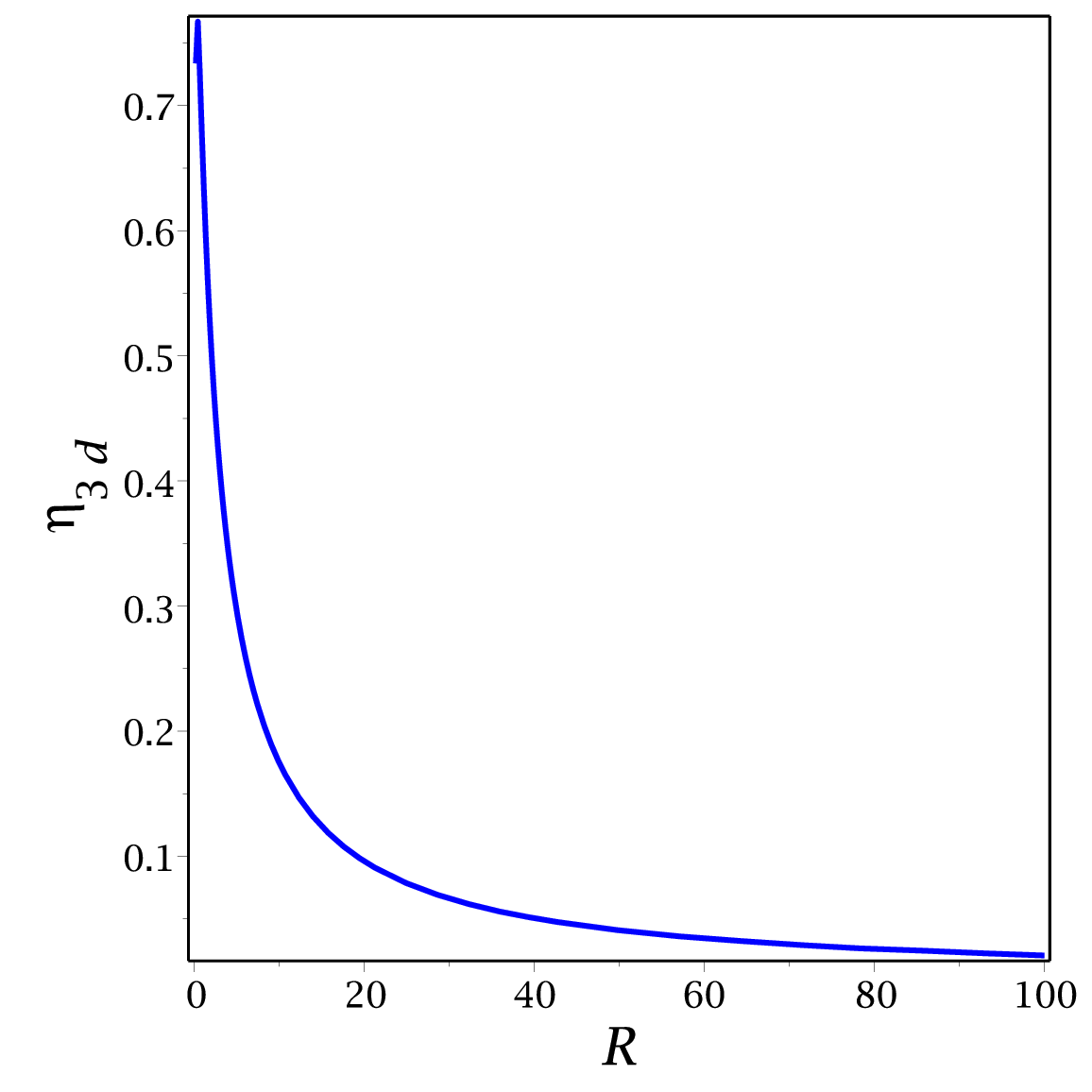}
\caption{Stability of $3D$ gravastars in massive gravity. We have chosen $%
\Lambda = -0.1$, $m = 1$, $c_1 = 0.1$, $C=0 .1$, $R = 1$, $A = 0$ and $m_0 =
0.01$.}
\label{eta3D}
\end{figure}

It is very useful to understand the stability of gravastars by defining a
parameter $\eta $ as the ratio of the derivatives of $\psi $ and $\varphi $
as follows,
\begin{equation}
\eta =\frac{\psi ^{\prime }(r)}{\varphi ^{\prime }(r)}.
\end{equation}
The stability regions can be explored by analyzing the behavior of $\eta$ as
a function of $r=R$. This parameter indicates the squared speed of sound
satisfying $0\leq\eta \leq 1$ \cite{59f}. It is possible, however, this
limitation is not met on the surface layer when testing the stability of the
gravastar \cite{120s,121s}. We have investigated the stable gravastars with
specific choices of parameters involved. Fig. \ref{eta3D} describes the
stability of $3D$ gravastar structures in massive gravity.

\subsection{Some Features of Intermediate Thin Shell of $3D$ Gravastars}

\label{the physical parameters}

This section aims to examine the impact of massive parameter on $3D$
gravastar's physical properties in the presence of massive gravity. In this
context, we examine the proper length of the thin shell and the energy of
the relativistic structure of $3D$ gravastars in massive gravity. Then, we
will calculate the entropy of the thin shell of $3D$ gravastars in this
theory of gravity, too. Also, we will present our results through diagrams.

\subsubsection{Proper Length of the Thin Shell}

\begin{figure}[tbp]
\centering
\includegraphics[scale=0.3]{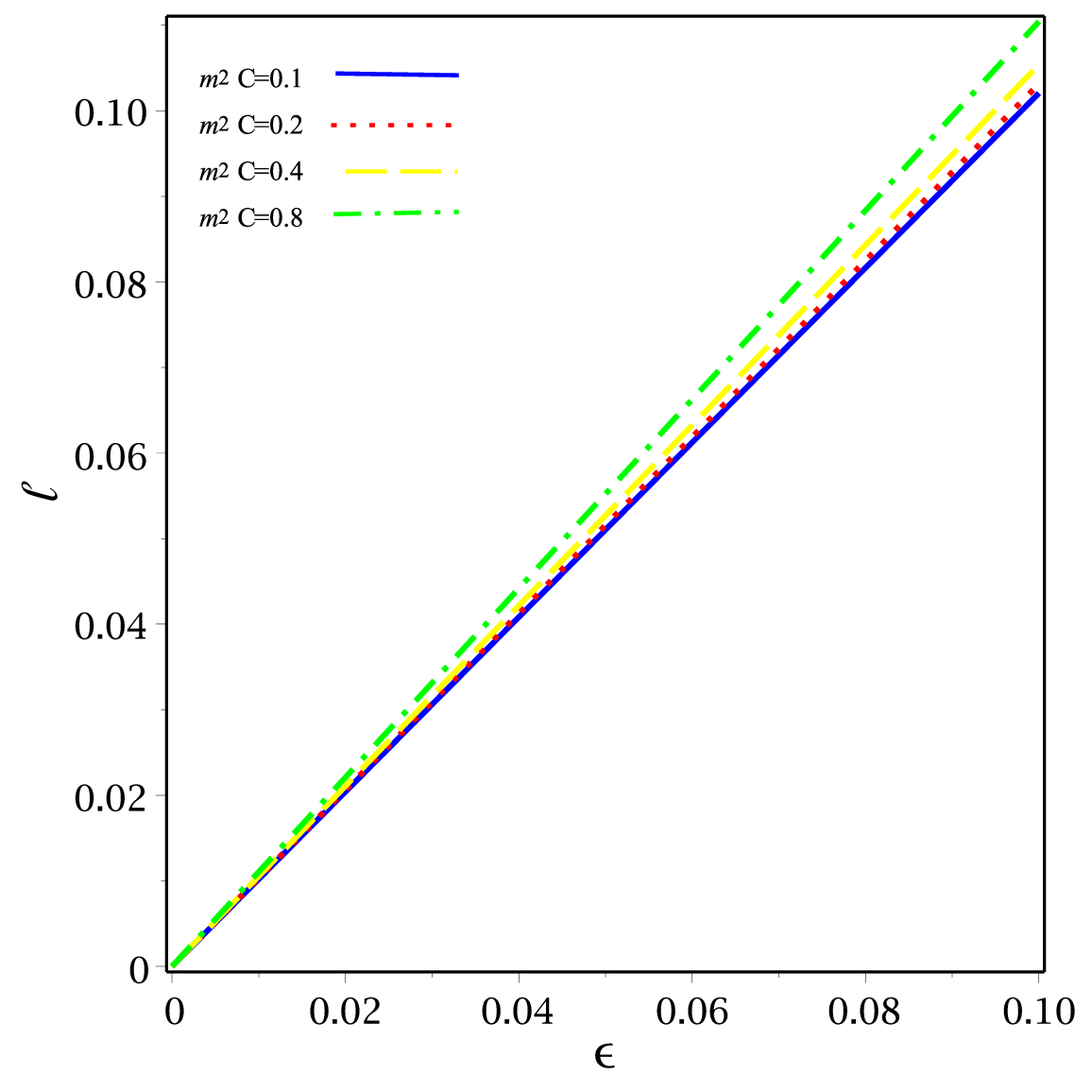}
\caption{Proper length $l$ of the shell vs thickness of the shell $\protect%
\epsilon $ for different values of $m^{2}C$. We have chosen $\Lambda =-0.1$,
$c_{1}=0.1$, $R=1$ and $B=1$.}
\label{lC}
\end{figure}

Since the radius of an interior region of gravastar is $r_{1}=R$, while the
radius of the exterior region is $r_{2}=R+\epsilon $, where $\epsilon $ is
the thickness of the intermediate thin shell which is assumed to be very
small (i.e., $\epsilon <<1$). So, the stiff perfect fluid propagates between
two boundaries of the thin shell region of the gravastar. Now, the proper
thickness between two surfaces can be described mathematically as \cite%
{29f,30f}
\begin{equation}
l=\int_{R}^{R+\epsilon }\sqrt{\frac{1}{g(r)}}dr,  \label{l}
\end{equation}%
whereas in the shell region, the expression of $g(r)$ is complicated, so the
analytic solution of the above expression is not possible. So, we will solve
it by numerical method and examine the behavior of massive parameters. let
us assume $\sqrt{\frac{1}{g(r)}}=\frac{dF(r)}{dr}$, based on the integral
above, we can write
\begin{eqnarray}
l &=&\int_{R}^{R+\epsilon }\frac{dF(r)}{dr}dr=[F(R+\epsilon )-F(R)]  \notag
\\
&\approx &\epsilon \frac{dF(r)}{dr}\mid _{R}=\epsilon \sqrt{\frac{1}{g(r)}}%
\mid _{R}.
\end{eqnarray}%
Since $\epsilon <<1$, so $O(\epsilon ^{2})\approx 0$. Therefore in the above
manipulation, we considere only the first-order term of $\epsilon $. Thus
for this approximation, the proper length will be
\begin{equation}
l\approx \frac{\epsilon }{\sqrt{B+2\Lambda R^{2}-2m^{2}c_{1}CR}}.
\end{equation}%
The above result shows that the proper length of the thin shell of $3D$
gravastar in massive gravity is proportional to the thickness $\epsilon $ of
the shell. We observe that the proper length of the thin shell depends on
the massive parameters as well. The behavior of the shell length against its
thickness for different values of $m^{2}c_{1}$ and $C$ is shown in Fig. \ref%
{l3D.e} and Fig. \ref{lC}, respectively.
\begin{figure}[tbp]
\centering
\includegraphics[scale=0.3]{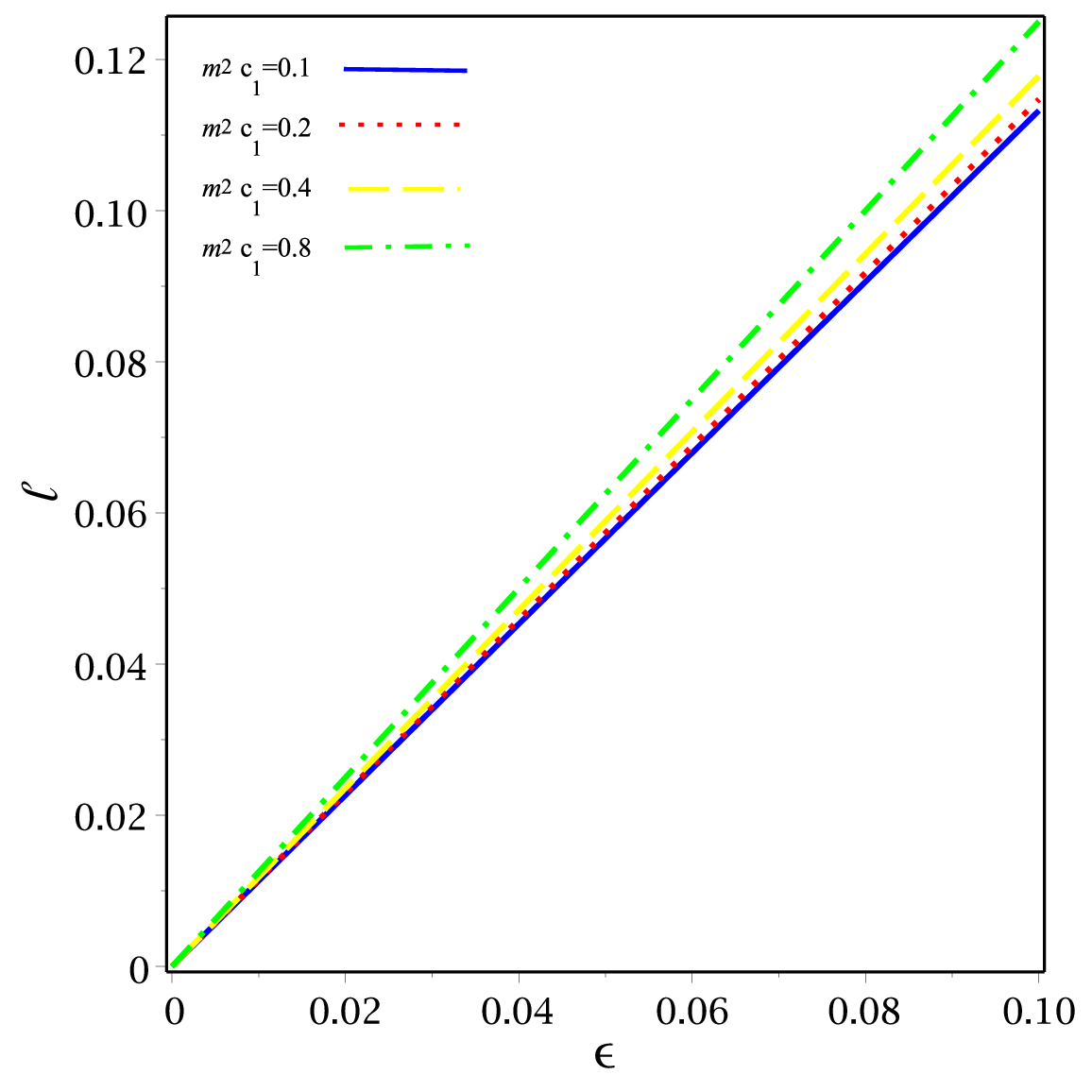}
\caption{Proper length $l$ of the shell vs thickness of the shell $\protect%
\epsilon $ for different values of $m^{2}c_{1}$. We have chosen $\Lambda
=-0.1$, $C=0.1$, $R=1$ and $B=1$.}
\label{l3D.e}
\end{figure}
Our results in the above figures show that there is a linear relationship
between the proper length and thickness of the shell, while the proper
length of the system tends to increase by increasing the corresponding $%
m^{2}c_{1}$, and $C$ values.

\subsubsection{Energy}

\begin{figure}[tbp]
\centering
\includegraphics[scale=0.3]{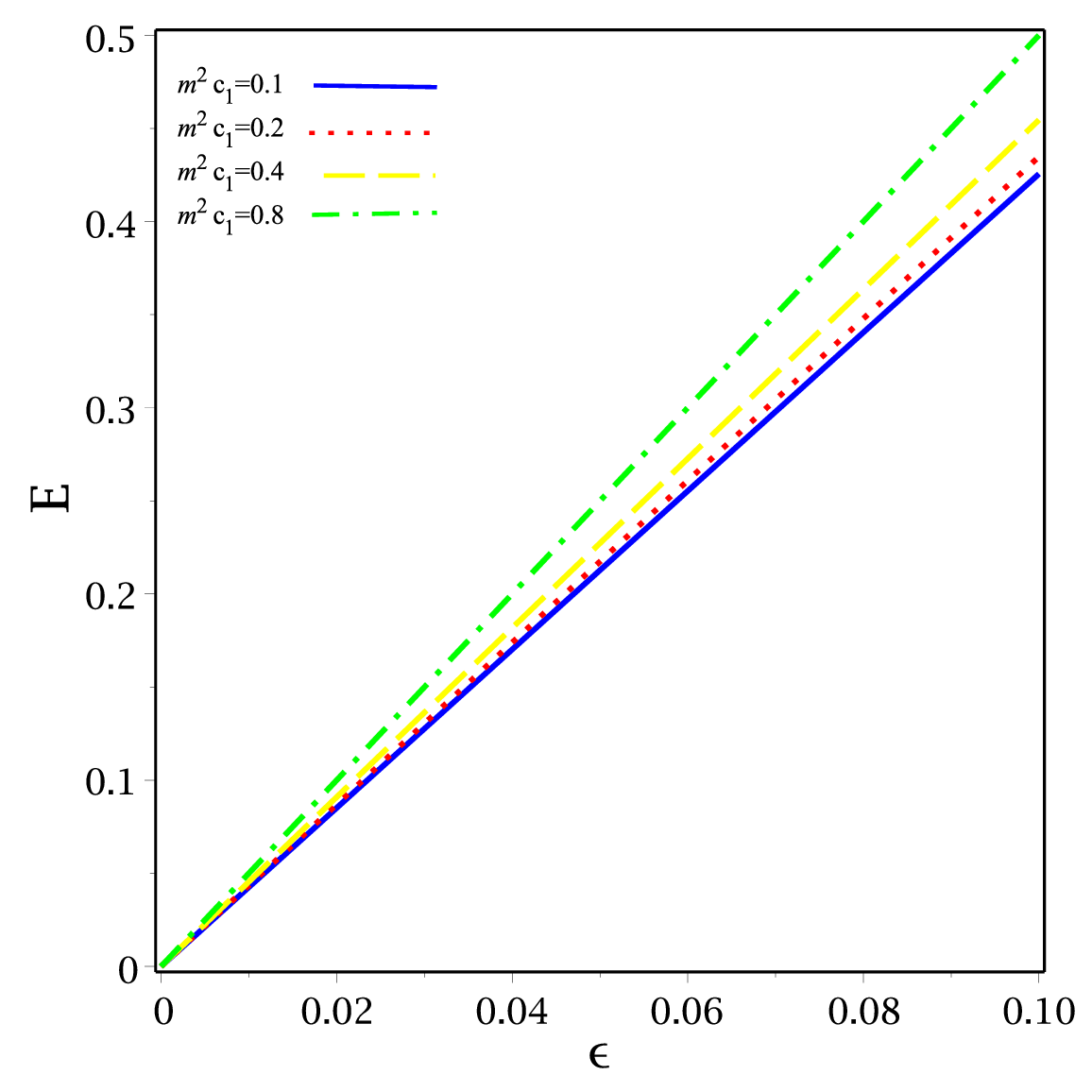}
\caption{Energy $E$ with thickness of the shell $\protect\epsilon $ for
different values of $m^2c_1$. We have chosen $\Lambda = -0.1$, $C=0 .1$, $R
= 1$ and $P_0= 1$.}
\label{E3D.e}
\end{figure}

\begin{figure}[tbp]
\centering
\includegraphics[scale=0.3]{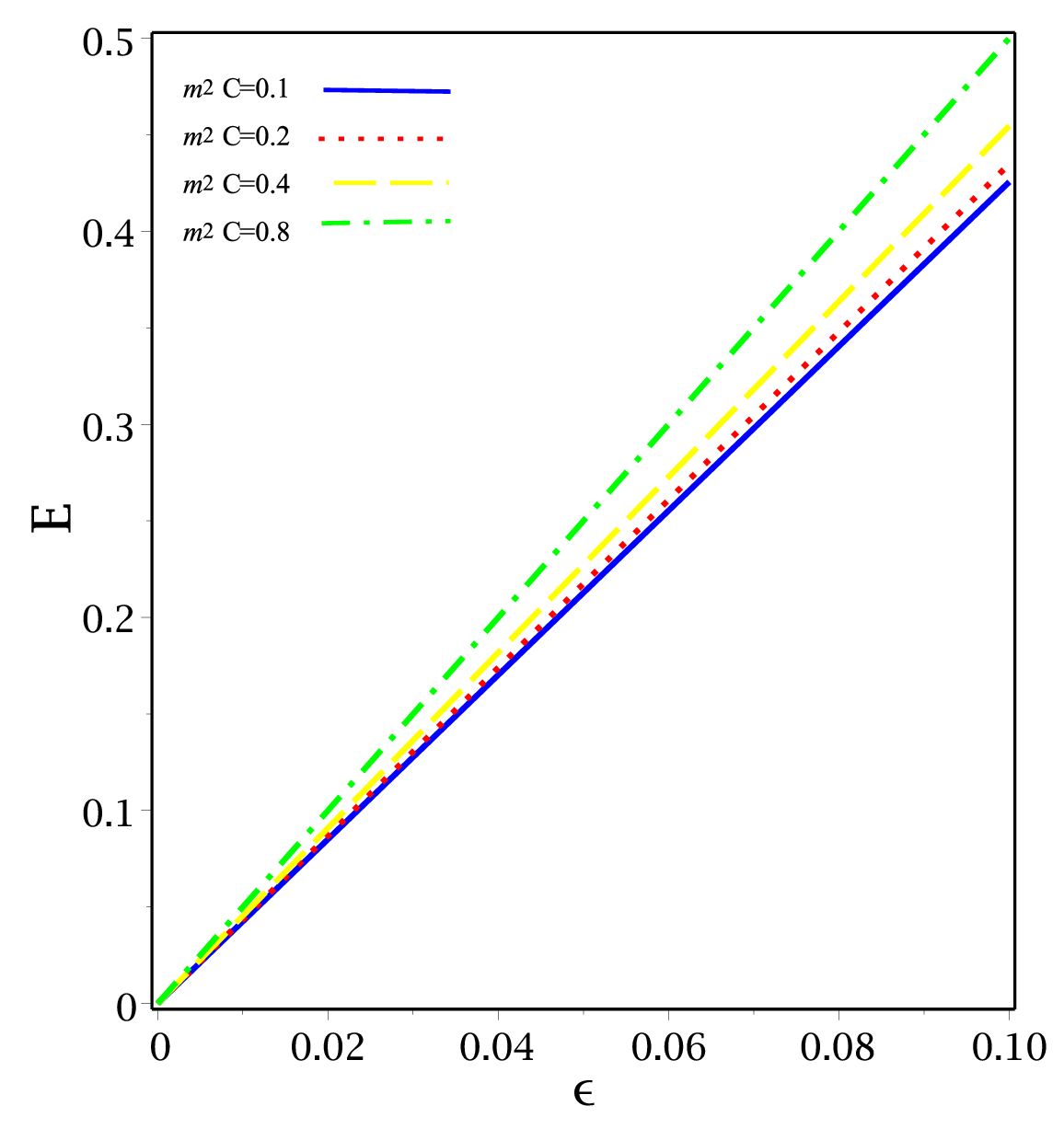}
\caption{Energy $E$ vs thickness of the shell $\protect\epsilon$ for
different values of $m^2C$. We have chosen $\Lambda = -0.1$, $c_1=0 .1$, $R
= 1$ and $P_0 = 1$.}
\label{EC}
\end{figure}

The energy content within the shell region of $3D$ gravastar is given as
\cite{29f,30f}
\begin{equation}
E=2\pi \int_{R}^{R+\epsilon }\rho rdr.
\end{equation}
By expanding $F(R+\epsilon )$ binomially about $R$ and taking first order of
$\epsilon $, we get
\begin{equation}
E\approx \frac{2\pi \epsilon p_{0}R^{2}}{2\Lambda R-m^{2}c_{1}C}.
\end{equation}%
The behavior of the shell energy against its thickness for different values
of $m^{2}c_{1}$ and $C$ is shown in Fig. \ref{E3D.e} and Fig. \ref{EC},
respectively.
The results in Fig. \ref{E3D.e} and Fig. \ref{EC} reveal the same behavior
for the shell energy. In other words, there is a linear relationship between
the energy and the thickness of the shell. In addition, this energy of the
system increases by increasing $m^{2}c_{1}$, and $C$, similar to the proper
length of the thin shell.

\subsubsection{Entropy}

\begin{figure}[tbp]
\centering
\includegraphics[scale=0.3]{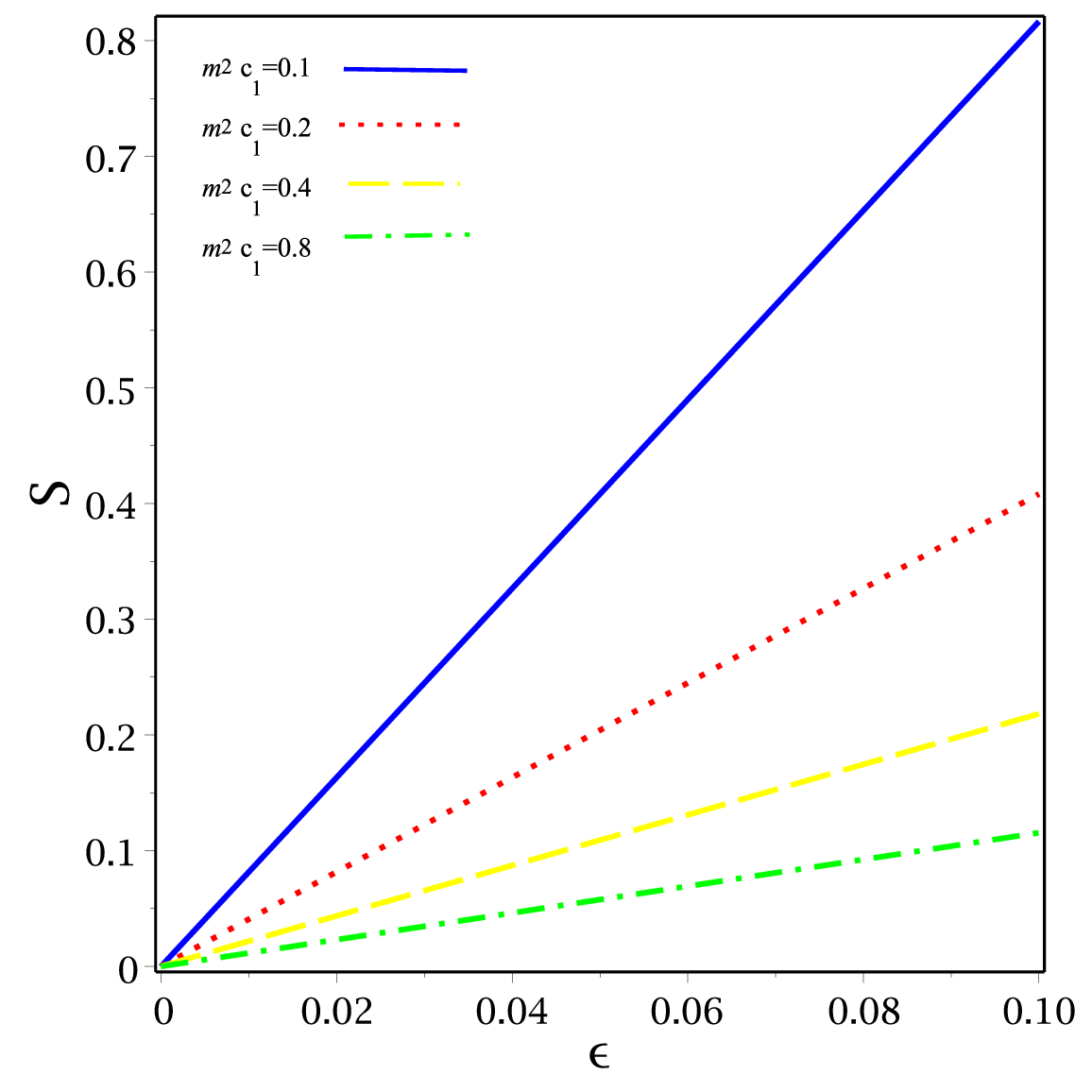}
\caption{Entropy $S$ vs thickness of the shell $\protect\epsilon $ for
different values of $m^{2}c_{1}$. We have chosen $\Lambda =-0.1$, $C=1$, $R=1
$, $P_{0}=1$ and $B=1$.}
\label{S3D.e}
\end{figure}

\begin{figure}[tbp]
\centering
\includegraphics[scale=0.3]{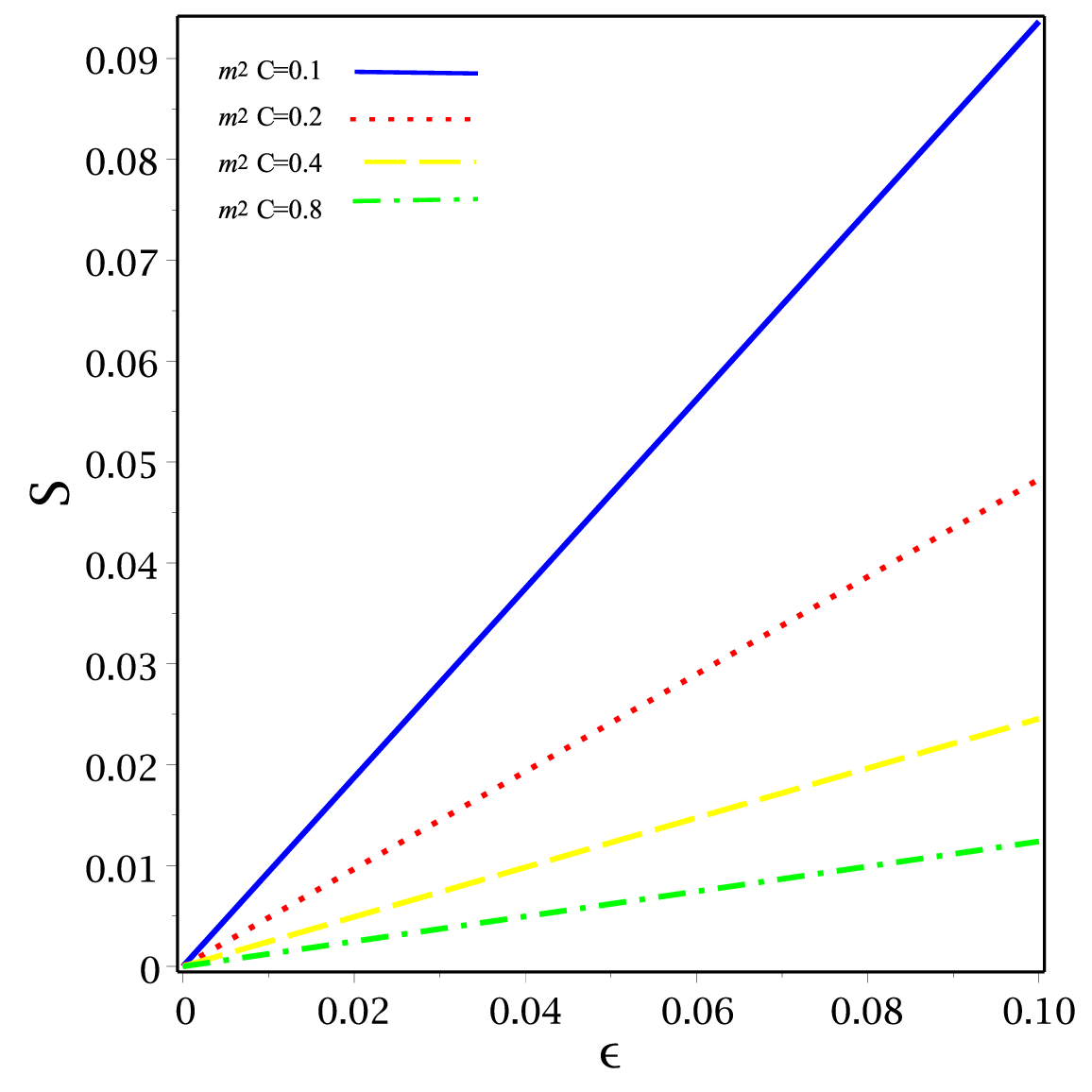}
\caption{Entropy $S$ vs thickness of the shell $\protect\epsilon $ for
different values of $m^{2}C$. We have chosen $\Lambda =-0.1$, $c_{1}=1$, $R=1
$, $P_{0}=1$ and $B=1$.}
\label{SC}
\end{figure}

Mazur and Mottola have shown that the entropy density in the interior region
of the gravastar is zero \cite{29f,30f}. To calculate the entropy relation
for the shell of $3D$ gravastar, we need to use the following equation \cite%
{29f,30f},
\begin{equation}
S=\int_{R}^{R+\epsilon }4\pi r^{2}s(r)\sqrt{\frac{1}{g(r)}}dr,  \label{s1}
\end{equation}%
where $s(r)$, describes the entropy density corresponding to a specific
temperature $T(r)$, is given by
\begin{equation}
s(r)=\frac{\alpha ^{2}k_{B}^{2}T(r)}{4\pi \hbar ^{2}}.  \label{s2}
\end{equation}%
Here $\alpha ^{2}$ is a dimensionless constant, due to the fact that we are
using Planck units $(K_{B}=\hbar =1)$ in our computation. Using Eqs. (\ref%
{s1}) and (\ref{s2}), the entropy inside a thin shell of $3D$ gravastar in
massive gravity can be written as
\begin{equation}
S=\frac{\alpha k_{B}\sqrt{2\pi p_{0}}\int_{R}^{R+\epsilon }\frac{rdr}{\sqrt{%
(B+2\Lambda r^{2}-2m^{2}c_{1}Cr)(2\Lambda -\frac{m^{2}c_{1}C}{r})}}}{\hbar }.
\end{equation}%
By expanding $F(R+\epsilon )$ binomially about $R$ and taking the first
order of $\epsilon $, we get
\begin{equation}
S\approx \frac{\alpha k_{B}\sqrt{2\pi p_{0}}\epsilon R}{\hbar \sqrt{%
(B+2\Lambda R^{2}-2m^{2}c_{1}CR)(2\Lambda -\frac{m^{2}c_{1}C}{R})}}.
\end{equation}%
The behavior of shell entropy against its thickness for different values of $%
m^{2}c_{1}$ and $C$ is shown in Fig. \ref{S3D.e} and Fig. \ref{SC},
respectively.
Fig. \ref{S3D.e} and Fig. \ref{SC} show the linear relationship between
entropy and thickness of the shell of $3D$ gravastars. Also, the entropy of
the system decreases by increasing $m^{2}c_{1}$ and $C$.

It is notable that the mass of graviton, one of the physical
principles of massive gravity, causes such a sensitive effect on
astrophysical consequences such as proper length, energy, and entropy.

\section{CONCLUSION}

In this work, we have investigated a new model of $3D$ gravastar with an
isotropic matter distribution AdS spacetime in massive gravity. Gravastars
are the same gravitational vacuum stars that define a new idea in the
gravitational system. Gravastar consists of three regions, the first is the
inner region, the second is the middle thin shell with a thickness of $%
\epsilon $, and the third is the outer region. Each of these regions is
described and investigated by a specific EoS. We found a set of
singularity-free solutions of gravastars and hence interesting results that
can be viewed as alternatives to BTZ black holes in massive gravity. 
In the interior region, we observed that the spacetime is the free
singularity. The physical parameters, such as density, pressure, and
gravitational mass in no way are dependent on the massive parameters ($m^{2}c_{1}$, and $C$) in the interior region, but the
quantities $g(r)$ and $f(r)$ depend on the massive
parameters. The exterior region with EOS $p=\rho =0$ is defined by
the static BTZ black hole in massive gravity, which can be seen in Eq. (\ref%
{E2}). At the junction interface, the interior region joins with the
exterior region with smooth matching at $r=R$. We derived some
aspects like surface energy density, surface pressure, EoS, and stability.
The EoS parameter depends upon the massive parameters, mass, and radius of
the metric. Fig. \ref{eta3D} describes the stability of $3D$
gravastar structures in massive gravity. It was not easy to find the exact
solution in the shell region with EoS, $p=\rho $. For this purpose,
we used the thin shell approximation as $0<g(r)\equiv h<<1$ to
extract the proper length, energy, and entropy for the shell region. Figs. %
\ref{l3D.e}, and \ref{lC} are plotted between the proper length of the shell
and the thickness of the shell. These figures indicate the linear
relationship between the proper length, and thickness of the shell, while
the proper length of the system increases by increasing $m^{2}c_{1}$, and $C$. Our results in Figs. \ref{E3D.e}, and \ref{EC} reveal
that the energy and thickness of the shell are directly proportional to
each other, and also, the energy of the system increases by increasing the
value of $m^{2}c_{1}$ and $C$. To see the role of
entropy, thickness, and massive parameters, we drew Fig. \ref{S3D.e}, and
Fig. \ref{SC}. These figures indicate that there is a linear relationship
between entropy and the thickness of the shell as well. In addition, the
entropy of the system decreases by increasing $m^{2}c_{1}$ and $C$.

\label{VI}

\label{section.conc}
\acknowledgements{We would like to thank the referee for the good comments and advice that improved this paper. H. Barzegar and M. Bigdel wish to
thank University of Zanjan research council. G. H. Bordbar wishes to thank the Shiraz University Research Council. B. Eslam Panah thanks the University of
Mazandaran. The University of Mazandaran has supported the work of B. Eslam
Panah by title "Evolution of the masses of celestial compact objects in
various gravity".}


\end{document}